\documentclass[preprint]{revtex4-2}

\usepackage{graphicx}
\usepackage{dcolumn}
\usepackage{bm}
\usepackage{amsfonts}
\usepackage{amsmath}
\usepackage{braket}
\usepackage[utf8]{inputenc}
\usepackage{hyperref}
\begin{document}

\title{Bell-State Quantum Holography with Metasurfaces}

\author{Qinmiao Chen$^{1,+}$}
\author{Guangzhou Geng$^{2,+}$}
\author{Hong Liang$^{1}$}
\author{Wai Chun Wong$^{1,3}$}
\author{Tailin An$^{1}$}
\author{Randy Stefan Tanuwijaya$^{1}$}
\author{Junjie Li$^{2}$}
\email{jjli@iphy.ac.cn}
\author{Jensen Li$^{1,3}$}
\email{jensenli@ust.hk}

\affiliation{$^{1}$Department of Physics, The Hong Kong University of Science and Technology, Hong Kong, P. R. China.}
\affiliation{$^{2}$Beijing National Laboratory for Condensed Matter Physics, Institute of Physics, Chinese Academy of Sciences, Beijing, P. R. China.}
\affiliation{$^{3}$Department of Engineering, University of Exeter, Exeter, United Kingdom.}
\affiliation{$^{+}$These authors contributed equally to this research.}

\begin{abstract}
Metasurfaces composed of subwavelength nanostructures enable simultaneous control of polarization and wavefront, greatly enhancing holographic information capacity. Building on this capability, we extend holography into the quantum domain by experimentally realizing Bell-state holograms—distinct holographic images encoded in polarization-entangled Bell states of photon pairs. A polarization-multiplexed dielectric metasurface generates spatial modes conditioned on both input and output polarizations, entangling the holographic pattern with the two-photon state. To characterize these quantum holograms, we further develop quantum hologram tomography, reconstructing the full density matrix of the holographic state pixel by pixel. The reconstructed density-matrix hologram reveals tailor-made holographic symbols attached to individual Bell states through the metasurface, with contrast built up among the different Bell components as theory shows. This framework unifies metasurface photonics with quantum-state reconstruction and provides a scalable route toward high-dimensional quantum communication, encryption and information processing based on holographically encoded quantum light.
\end{abstract}

\maketitle

\section{Introduction}

Holography, first introduced by Dennis Gabor in 1948 to record both amplitude and phase information of optical waves~\cite{Gabor1948Nature}, has evolved from optical interference to computer-generated and metasurface implementations, enabling precise wavefront control and multidimensional information encoding~\cite{Leith1962JOSA,Jiang2019AOP}. 
As a core technique for light-field modulation, holography now underpins diverse fields including optical displays, interferometric metrology, microscopy, data storage, and information security~\cite{Blanche2021LAM,Stetson1966JOSA,Kuznetsova2007OE,Heanue1994Science,Refregier1995OL}. 

Conventional holographic devices such as spatial light modulators (SLMs), diffractive optical elements (DOEs), and digital micromirror devices (DMDs) are limited by large form factors, restricted resolution, and low efficiency~\cite{Huang2012AO,Sun2013Nature,Kreis2001OE}. 
The emergence of metasurfaces—ultrathin arrays of subwavelength nanostructures—provides a transformative route to overcome these bottlenecks by granting precise control of amplitude, phase, and polarization at the subwavelength scale. 
Metasurfaces have demonstrated exceptional performance in applications such as achromatic metalenses~\cite{Chen2018NatNano,Wang2021NatCommun}, miniaturized microscopes~\cite{Kwon2020NatPhoton}, laser pulse shaping~\cite{Divitt2019Science}, and structured-beam generation~\cite{Ming2022AdvMater,Kim2022NatCommun,Chen2024NatNano}. 
Compared with conventional holography, metasurface-based holography (meta-holography) leverages deep-subwavelength pixelation for high-precision light-field control~\cite{Chen2024NatNano}. 
The resulting high resolution~\cite{Ni2013NatCommun,Huang2013NatCommun}, high diffraction efficiency~\cite{Zheng2015NatNano,Wang2016Optica}, and wide viewing angle~\cite{Huang2013NatCommun,Xiong2025AOM} firmly establish meta-holography as a leading platform for compact, high-quality, and multifunctional holographic devices. 
Beyond structural miniaturization, expanding the dimensionality and security of holographic information is key to enhancing the capacity and robustness of optical information processing and encryption.

Integrating quantum light with holography further extends these capabilities. 
Quantum light sources exploit photon entanglement, offering unique advantages for quantum communication and imaging. 
Strong correlations between entangled photon pairs enable nonlocal and interaction-free imaging, leading to mechanisms such as ghost imaging~\cite{Lemos2014Nature,Yang2023npjQI}. 
These quantum-imaging schemes achieve high sensitivity~\cite{Brida2010NatPhoton}, reduced photodamage~\cite{Zhang2024SciAdv}, and enhanced resolution~\cite{Tenne2019NatPhoton}. 
Moreover, combining entangled photons with structured light allows hybrid or hyper-entanglement between polarization and spatial modes~\cite{Liang2023CommunPhys,Ornelas2024NatPhoton}, which underpins high-dimensional quantum communication~\cite{Forbes2019AVSQS}. 
Building on these advances, quantum holography based on entangled photons exhibits exceptional noise resistance and supports ultra-high-dimensional spatially entangled modes~\cite{Defienne2021NatPhys}, which have been exploited for super-resolution imaging~\cite{Defienne2022NatCommun}. 
The multidimensional light-field modulation offered by metasurfaces opens new opportunities for manipulating entangled photons~\cite{Yousef2025Science}. 
Recent demonstrations—including multi-channel quantum holography with single-photon sources~\cite{Yang2022PhotonRes}, remote holographic pattern switching using polarization-entangled photon pairs~\cite{Fan2024APN}, and quantum-eraser-style holographic control~\cite{Liang2025AdvPhoton}—highlight the potential of metasurface-enabled quantum holography as a platform for high-dimensional quantum information encoding~\cite{Cozzolino2019AQT}.

The four Bell states, representing maximally entangled two-photon polarization states, form the foundation for numerous protocols in quantum communication~\cite{Bennett1993PRL,Ursin2007NatPhys}, quantum cryptography~\cite{Ekert1991PRL}, and quantum computation~\cite{Zurita2025Quantum}. A series of important advances have been made in metasurface-based generation and measurement of Bell states~\cite{Jia2025SciAdv,Shi2025AOM,Gao2023Nanophotonics}. Here, we introduce and experimentally demonstrate density-matrix holography, in which polarization-entangled Bell states serve as the encoding basis for generating and reconstructing distinct quantum holographic patterns. 
Central to our approach is a polarization-multiplexed dielectric metasurface that couples polarization and spatial degrees of freedom, producing a hybrid entangled state between the two photons and the holographic field. 
This interaction embeds two-photon entanglement directly within the hologram, linking spatially encoded information to the underlying quantum correlations. 
To fully characterize this hybrid state, we employ the density-matrix formalism, which captures both statistical mixtures and quantum coherences~\cite{Peres1996PRL}. 
We develop a quantum hologram tomography technique to reconstruct the two-photon holographic state at the density-matrix level, achieving the first experimental realization of a density-matrix hologram. 
By projecting the reconstructed density matrix onto the four Bell states—constituting a complete basis of the two-photon polarization Hilbert space~\cite{Dada2011NatPhys,Brukner2004PRL}—we reveal distinct, non-overlapping holographic patterns associated with each Bell component. 
This framework unites metasurface photonics with quantum-state reconstruction, significantly expanding the dimensionality of holographic encoding and offering a scalable route toward high-capacity and high-security quantum information processing. 
Furthermore, the information encoded in the reconstructed density matrix suggests promising applications in high-density quantum data storage and quantum computation~\cite{Viamontes2005QIC,Patel2023JIIS}.

\section{Working Principle of Bell-state Holograms}

\begin{figure}[t]
  \centering
  \includegraphics[width=\linewidth]{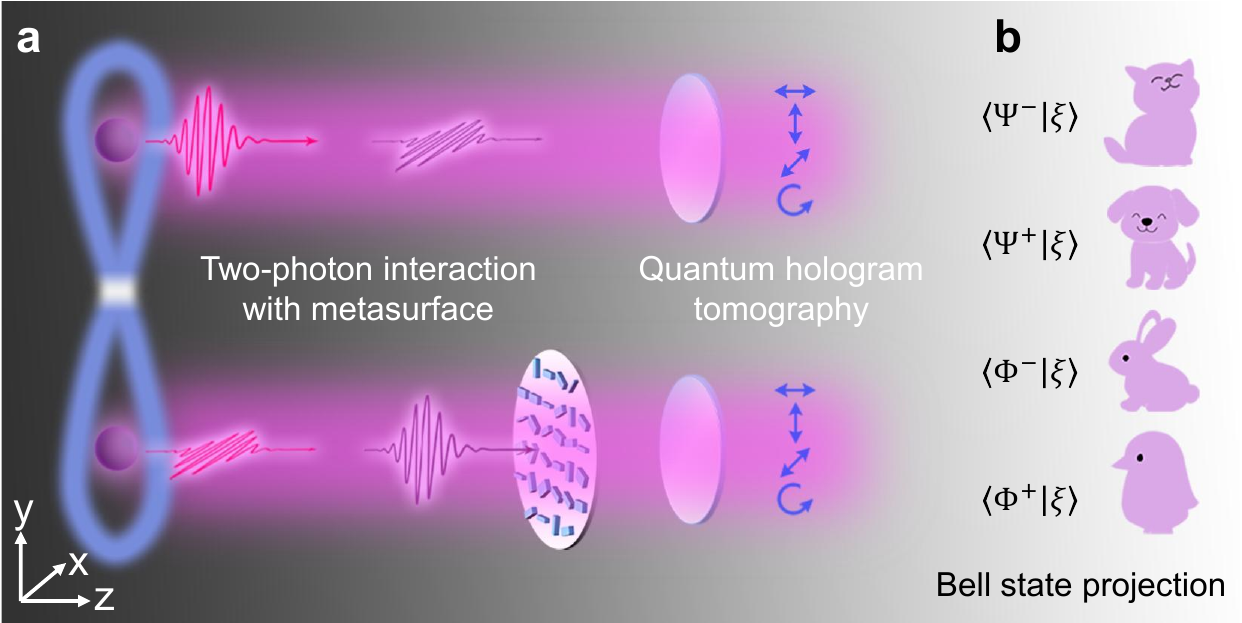}
  \caption{Schematic of Bell-state holograms. Polarization-entangled photon pairs interact with a polarization-multiplexed metasurface. In the general framework, distinct holographic symbols can be attached to all four polarization Bell states, establishing Bell-state holograms. In our specific implementation, three symbols (``='', ``$+$'', ``$\times$'') are attached to three Bell states, while the fourth state is left blank. The metasurface action determines the relative weights of the holographic channels through its polarization conversion properties.}
  \label{fig:workingprinciple}
\end{figure}

Figure~\ref{fig:workingprinciple} illustrates the generic concept of Bell-state holograms. A source generates polarization-entangled photon pairs, one of which passes through a polarization-multiplexed metasurface. The metasurface is engineered so that each polarization Bell state of the two-photon system can be associated with a distinct holographic symbol in the output arm. In the most general framework, this allows all four Bell states to carry independent holographic information through entanglement. In practice, however, one may also choose whether or not to attach a holographic image to each Bell state, so that specific device implementations realize only a subset of these channels. In the present design, three holographic symbols, ``='', ``$\times$'', and ``$+$'', are intentionally assigned to three Bell states, while the remaining Bell state is left without any attached holographic images.

The target two-photon hologram state can be written in the most general form as
\begin{equation}
\label{eq:twoPhotonHologram}
\ket{\xi} \;=\; a_1\, \ket{\Phi^-}\!\otimes\!\ket{=}_b
      \;+\; a_2\, \ket{\Psi^+}\!\otimes\!\ket{\times}_b
      \;+\; a_3\, \ket{\Psi^-}\!\otimes\!\ket{+}_b \, ,
\end{equation}
where $a_1,a_2,a_3$ are complex weighting amplitudes. The four polarization Bell states are
\[
\ket{\Psi^+} = \tfrac{1}{\sqrt{2}}\!\left(\ket{L}_a\ket{R}_b + \ket{R}_a\ket{L}_b\right), \quad
\ket{\Psi^-} = \tfrac{1}{\sqrt{2}}\!\left(\ket{L}_a\ket{R}_b - \ket{R}_a\ket{L}_b\right),
\]
\[
\ket{\Phi^+} = \tfrac{1}{\sqrt{2}}\!\left(\ket{L}_a\ket{L}_b + \ket{R}_a\ket{R}_b\right), \quad
\ket{\Phi^-} = \tfrac{1}{\sqrt{2}}\!\left(\ket{L}_a\ket{L}_b - \ket{R}_a\ket{R}_b\right),
\]
where $L$ and $R$ denote left- and right-handed circular polarizations, and subscripts $a$ and $b$ label the two photons (separate spatial modes).

To realize Eq.~\eqref{eq:twoPhotonHologram}, we employ a metasurface with a single layer of nanostructures that implements a polarization-multiplexing operator acting on photon $b$:
\begin{equation}
\label{eq:M_operator}
\begin{aligned}
\hat{M}\ket{L}_b &= \alpha\, \ket{L,=}_b \;+\; \beta\!\left(\ket{R,+}_b + \ket{R,\times}_b\right), \\
\hat{M}\ket{R}_b &= \alpha\, \ket{R,=}_b \;+\; \beta\!\left(\ket{L,+}_b - \ket{L,\times}_b\right).
\end{aligned}
\end{equation}
Here, $\alpha$ and $\beta$ are device-controlled amplitudes determined by the metasurface nanostructure. As the incident state, we consider a polarization-entangled pair such as
\begin{equation}
\label{eq:inputState}
\ket{\Phi^-}_{ab} = \tfrac{1}{\sqrt{2}}\!\left(\ket{L}_a\ket{L}_b - \ket{R}_a\ket{R}_b\right).
\end{equation}

Applying $\hat{M}$ to photon $b$ of this entangled state and regrouping terms using the Bell-state definitions yields
\begin{equation}
\label{eq:output}
(\mathbb{I}_a \!\otimes\! \hat{M}_b)\ket{\Phi^-}_{ab}
= \alpha\, \ket{\Phi^-}\!\otimes\!\ket{=}_b
 \;+\; \beta\, \ket{\Psi^+}\!\otimes\!\ket{\times}_b
 \;+\; \beta\, \ket{\Psi^-}\!\otimes\!\ket{+}_b \, .
\end{equation}
Thus, in Eq.~\eqref{eq:twoPhotonHologram}, the coefficients are fixed by the metasurface action as $a_1=\alpha$, $a_2=\beta$, and $a_3=\beta$. More details are given for the derivation in \emph{Supplementary Note 1}. The co-polarized channels (LL, RR) generate the ``='' holographic image with weight $\alpha$, while the cross-polarized channels (LR, RL) generate the ``+'' and ``$\times$'' holographic images with equal weight $\beta$ and relative phases of $0$ and $\pi$. In deriving Eq.~\eqref{eq:M_operator}, we assumed a mirror symmetry of the metasurface along the propagation direction. This symmetry constrains the two cross-polarized channels to have equal weights, so that $a_2=a_3$. These details reflect the specifics of our implementation for simplicity and do not represent fundamental restrictions of the Bell-state hologram framework.

Moreover, while the mapping can be described in terms of Bell-state holograms, a full description of the system requires the density matrix of the two-photon state. The density-matrix representation captures not only the probabilities of each holographic image but also the quantum coherences between them. In this sense, the holographic patterns are encoded in the density matrix itself, motivating the broader concept of density-matrix holography and its reconstruction via quantum hologram tomography. For the present example, the diagonal element corresponding to the unused Bell state later appears dark in the reconstructed density-matrix images. 

\section{Metasurface Design}
\label{sec:metasurface}

\begin{figure}[t]
  \centering
  \includegraphics[width=\linewidth]{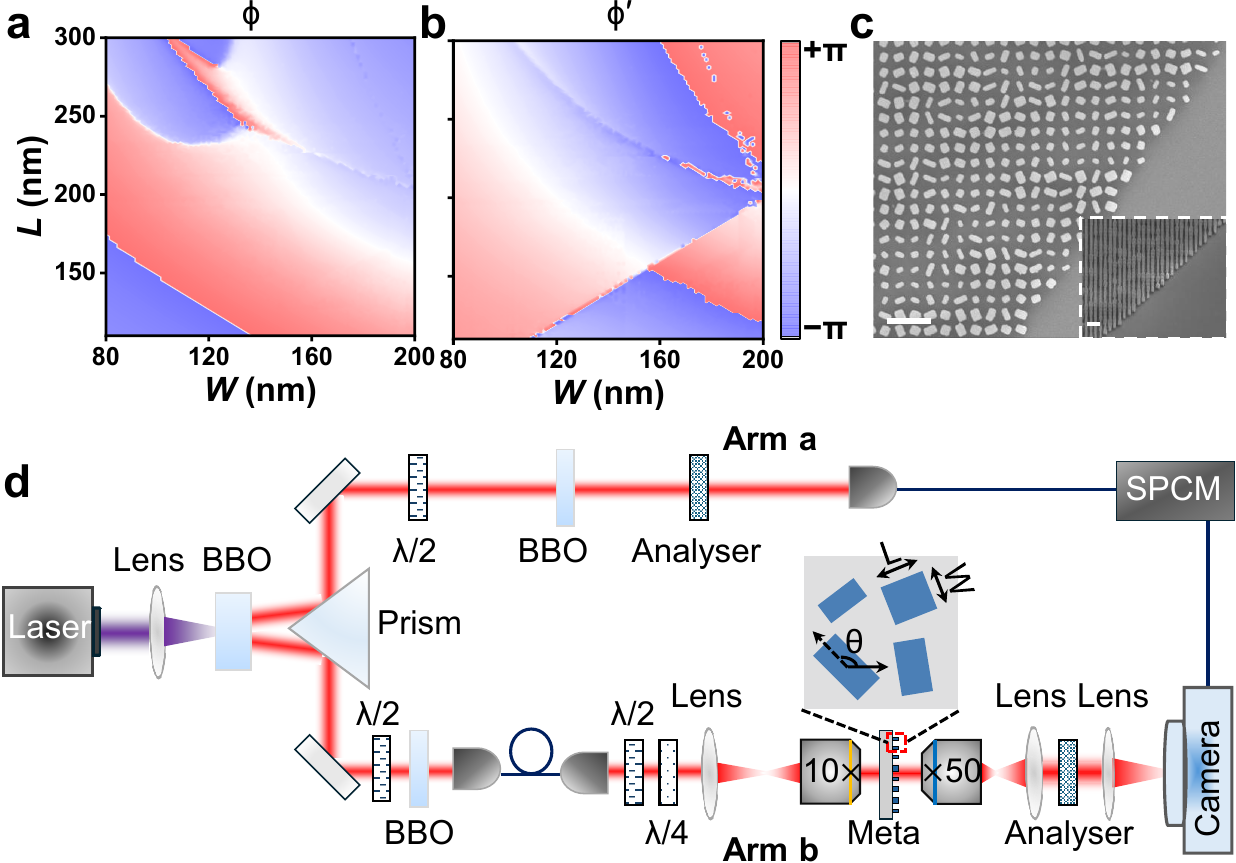}
  \caption{Metasurface design and fabrication. 
  (a) Phase map of the co-polarized propagation phase $\phi$ as a function of nanopillar length $L$ and width $W$ (with fixed height $H=500$~nm). 
  (b) Phase map of the cross-polarized propagation phase $\phi'$. 
  (c) SEM image of the fabricated metasurface (square lattice, period $P=350$~nm); scale bar: $1\,\mu$m. Inset: tilted view. 
  (d) Experimental setup for quantum hologram tomography (introduced in the next section). Inset: top-view schematic of a single-pillar unit cell defining $L$, $W$, and $\theta$.}
  \label{fig:metasurface}
\end{figure}

To realize the polarization-multiplexing operator $\hat{M}$ in Eq.~\eqref{eq:M_operator}, we design a dielectric metasurface that provides independent control of co- and cross-polarized transmission channels. The metasurface is a square lattice of unit cells with in-plane periodicity $P=350$~nm. Each unit cell contains a single amorphous silicon (a-Si) nanopillar on a SiO$_2$ substrate. The refractive indices used in the design are $n_{\mathrm{SiO_2}}=1.45$ and $n_{\alpha\text{-Si}}=3.44$ at the operating wavelength of 810~nm. A nanopillar is characterized by its in-plane length ($L$), width ($W$), out-of-plane height ($H$), and in-plane rotation angle ($\theta$) measured from the $x$-axis. In our implementation, the height is fixed at $H=500$~nm to ensure sufficient optical path length for full $2\pi$ phase coverage; the tunable parameters are $L$, $W$, and $\theta$. This geometry is illustrated in the inset of Fig.~\ref{fig:metasurface}(d), and the SEM image in Fig.~\ref{fig:metasurface}(c) shows the fabricated array with spatially varying in-plane dimensions and orientations.

To describe its optical response, we start from the Jones matrix representation. For $\theta=0$ (pillar long axis along $x$), the transmission along the principal axes is described by $t_{xx}=|t_{xx}|e^{i\phi_{xx}}$ and $t_{yy}=|t_{yy}|e^{i\phi_{yy}}$. Expressed in the circular-polarization (CP) basis $\{L,R\}$, the device-level Jones matrix we use throughout the paper is
\begin{equation}
\label{eq:JCP}
J_{\mathrm{CP}} =
\begin{bmatrix}
t_{LL} & t_{LR} \\
t_{RL} & t_{RR}
\end{bmatrix}
=
\begin{bmatrix}
|t|\, e^{i\phi} & |t'|\, e^{i(\phi' - 2\theta)} \\
|t'|\, e^{i(\phi' + 2\theta)} & |t|\, e^{i\phi}
\end{bmatrix}.
\end{equation}
Here, the diagonal terms describe co-polarized transmission with propagation phase $\phi$, while the off-diagonal terms describe cross-polarization conversion with propagation phase $\phi'$ and spin-dependent geometric phases $\pm 2\theta$. For completeness, the CP coefficients follow from the standard LP$\to$CP basis transform:
\[
t_{LL}=t_{RR}=\tfrac{1}{2}(t_{xx}+t_{yy}), \quad 
t_{LR}=\tfrac{1}{2}(t_{xx}-t_{yy})\,e^{-i2\theta}, \quad
t_{RL}=\tfrac{1}{2}(t_{xx}-t_{yy})\,e^{+i2\theta},
\]
so that $\phi=\arg(t_{LL})=\arg(t_{RR})$ and $\phi'=\arg(\sqrt{t_{LR}}\sqrt{t_{RL}})$ (see \emph{Supplementary Note 2} for details). To verify that the metasurface can supply the required phases, we numerically mapped the accessible ranges of $\phi$ and $\phi'$ by sweeping $L=110$–$300$~nm and $W=80$–$200$~nm (with $H=500$~nm). The resulting phase maps in Figs.~\ref{fig:metasurface}(a,b) exhibit full $2\pi$ coverage for both quantities, establishing a complete library (together with the amplitude maps in Fig. S1 in \emph{Supplementary Note 3}) of unit-cell responses from which the metasurface is assembled.

With these phase controls available, the metasurface enforces the relations required by $\hat{M}$ in Eq.~\eqref{eq:M_operator}: the co-polarized channels (LL, RR) contribute to the ``='' holographic image via $\phi$, while the cross-polarized channels (LR, RL) contribute to the ``$+$'' and ``$\times$'' holographic images via $\phi' \pm 2\theta$. The $\pi$ phase difference between the two holographic symbols "$+$" and "$\times$" in the two cross-polarization channels is implemented by an additional constraint in the Gerchberg-Saxton (G-S) algorithm in generating the near-field phase profiles, with more details given in \emph{Supplementary Note 4}. Then the joint control of these near field phases yields three coherent, independently addressable holographic fields, as required to realize the Bell-state hologram state in Eq.~\eqref{eq:twoPhotonHologram}. The fabricated device is shown in Fig.~\ref{fig:metasurface}(c), which confirms the square lattice ($P=350$~nm) and the targeted spatial variation of $(L,W,\theta)$ across the aperture, thereby providing the physical realization of the polarization-multiplexing operator $\hat{M}$.

\section{Quantum Hologram Tomography}
\label{sec:tomography}

\begin{figure}[t]
  \centering
  \includegraphics[width=0.8 \linewidth]{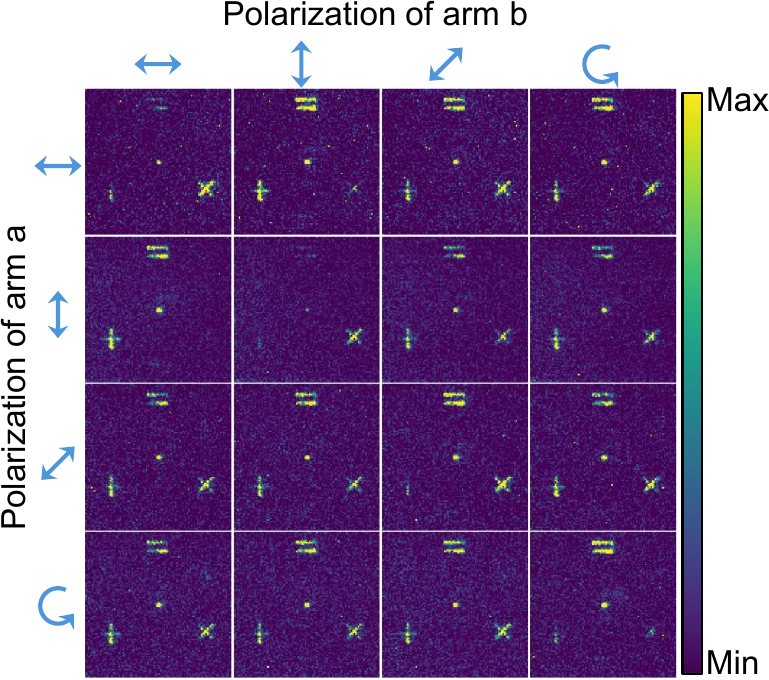}
  \caption{Measured polarization-projected holographic images for quantum hologram tomography. Sixteen holographic images were obtained under different polarization-projection combinations of the entangled photon pairs, recorded via coincidence detection between a single-photon counting module (arm $a$) and the SPAD camera (arm $b$). The vertical axis corresponds to the analyser settings in arm $a$, while the horizontal axis corresponds to those in arm $b$. Polarization states used were horizontal (H), vertical (V), diagonal (D), and left-circular (L).}
  \label{fig:expdata}
\end{figure}

The polarization-multiplexed metasurface was fabricated from a 500-nm-thick amorphous silicon film on a SiO$_2$ substrate using electron-beam lithography and dry etching (see Methods). As shown in Fig.~\ref{fig:metasurface}(c), scanning electron microscope (SEM) images reveal well-defined nanopillar arrays with rectangular cross-sections, precisely controlled lateral dimensions, and varying orientation angles. The clean vertical sidewalls confirm that both propagation and geometric phases are accurately implemented, ensuring that the fabricated device faithfully realizes the polarization-multiplexing operator $\hat{M}$ required for the Bell-state holograms.

The experimental setup is depicted in Fig.~\ref{fig:metasurface}(d). A continuous-wave diode laser at 405~nm pumps a $\beta$-barium borate (BBO) crystal to produce polarization-entangled photon pairs at 810~nm via type-II spontaneous parametric down-conversion (SPDC). The generated state can be expressed as
\(
\frac{1}{\sqrt{2}}\big(\ket{H}_a\ket{V}_b + \ket{V}_a\ket{H}_b\big),
\)
which is equivalent to Eq.~\eqref{eq:inputState} up to a local unitary transformation. The photon pairs are separated by a polarizing beam splitter (PBS). Photon $a$ is directed to a single-photon counting module (SPCM), which serves as a trigger for time-gated coincidence detection on a single-photon avalanche diode (SPAD) camera placed in arm $b$. Photon $b$ is transmitted through the metasurface, and its spatially resolved holographic response is recorded by the SPAD camera. Polarization analysers are placed in both arms to project the photons onto chosen polarization states prior to detection.

To capture the full information of the entangled two-photon hologram state, it is not sufficient to measure intensity patterns alone. The hologram is encoded both in the measurement probabilities of different outcomes and in the quantum coherences between them. A complete characterization therefore requires reconstruction of the two-photon density matrix. A general two-qubit density matrix contains 15 independent real parameters, corresponding to populations (diagonal elements) and coherences (off-diagonal elements). Including normalization, a minimum of 16 independent measurement settings is required for complete reconstruction. In standard two-photon tomography, these settings are obtained by projecting both photons onto four polarization states. In our experiment, we selected horizontal (H), vertical (V), diagonal (D), and left-circular (L) projections in each arm, yielding $4\times 4 = 16$ basis combinations. For each combination, coincidence counts were acquired pixel by pixel between photon $a$ and photon $b$, allowing us to reconstruct the spatially resolved density matrix. This pixel-by-pixel procedure constitutes the density-matrix hologram in this work.

The experimental results are presented in Fig.~\ref{fig:expdata}. Sixteen holographic images were recorded, each corresponding to one polarization-projection combination of photons in arms $a$ and $b$. The vertical axis of Fig.~\ref{fig:expdata} labels the analyser settings for arm $a$, while the horizontal axis labels those for arm $b$. The measured intensity distributions clearly demonstrate how the holographic sub-patterns appear, disappear on the joint polarization projections of the entangled photons. This behaviour follows the selection rules derived from the Bell-state hologram state (Eq.~\eqref{eq:twoPhotonHologram}): for example, the HH and VV projections isolate the ``$\times$'' holographic sub-pattern, HV (or VH) yield a superposition of ``='' and ``$+$'' sub-patterns, the DD projection selects both ``='' and ``$\times$'' sub-patterns, and the LL projection selects only the ``='' sub-pattern. Mixed-basis settings such as HD, HL, or VD display all three sub-patterns simultaneously. These observations confirm the expected correlations between analyzer settings and hologram channels, providing direct confirmation for the design principle of the quantum hologram. The data show excellent agreement with numerical predictions (Fig.~S5), with residual noise mainly attributed to dark counts of the SPAD camera.

\section{Density-Matrix Reconstruction of Bell-State Holograms}

\begin{figure}[t]
  \centering
  \includegraphics[width=\linewidth]{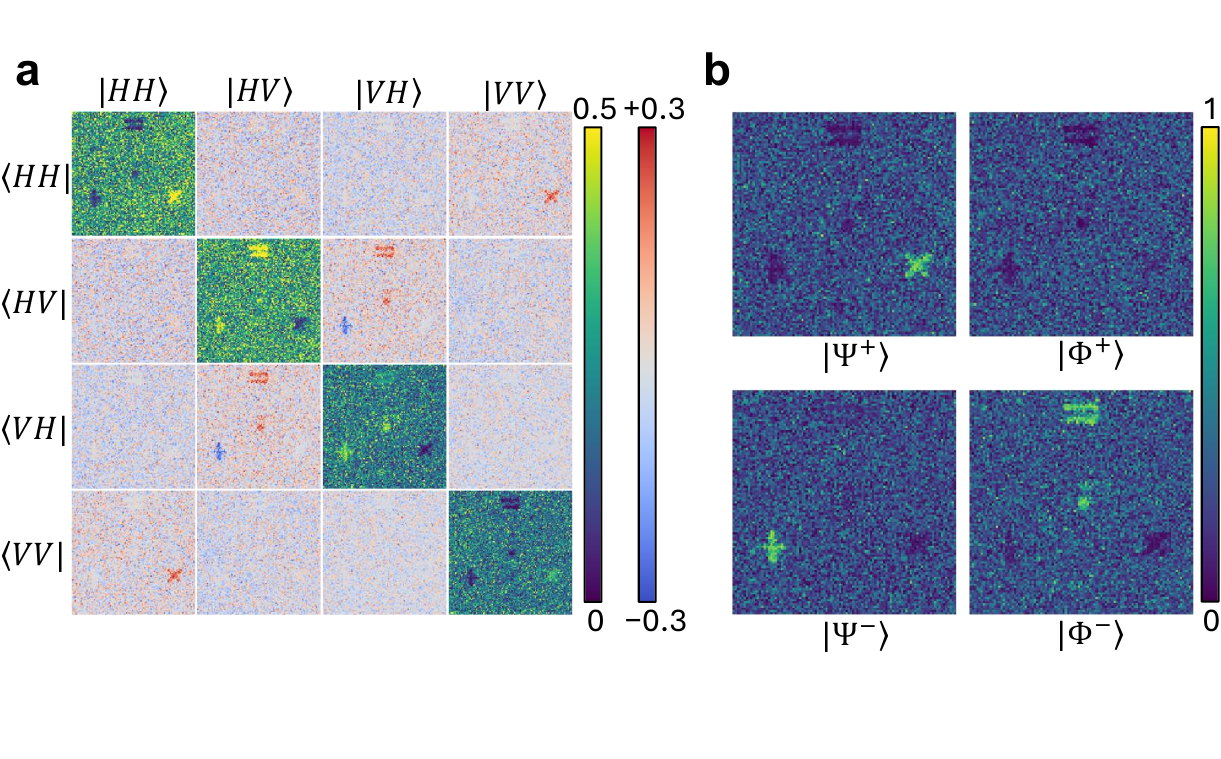}
  \caption{Density-matrix hologram and Bell-basis readout. 
  \textbf{a}, Real part of the reconstructed density-matrix hologram $\Re[\rho(x,y)]$ in the linear basis. Each matrix element is a 2D image rather than a scalar, encoding spatial holographic information. 
  \textbf{b}, Bell-basis projections $\langle \psi_B | \rho(x,y) | \psi_B \rangle$ reveal the symbol-selective encoding: ``$\times$'' for $|\Psi^+\rangle$, ``$+$'' for $|\Psi^-\rangle$, ``$=$'' for $|\Phi^-\rangle$, and blank for $|\Phi^+\rangle$, consistent with Eq.~\eqref{eq:twoPhotonHologram}. Purple–yellow and blue–red color bars are used for the positive-only diagonal and off-diagonal elements, respectively.}
  \label{fig:densitymatrix}
\end{figure}

Using the 16 polarization-projected holographic images from quantum hologram tomography (Fig.~\ref{fig:expdata}), we reconstruct the pixel-resolved two-photon density matrix by maximum-likelihood estimation following standard two-qubit tomography practice~\cite{James2001PRA}. The experimental counts for these sixteen settings are obtained as the difference between measurements with the pump laser on (signal + background) and off (background only). The resulting set of background-subtracted intensities serves as the input data $p_k$ for reconstructing a two-photon density matrix $\rho_i$ corresponding to each pixel $i$. The reconstruction is carried out by minimizing a Pearson-type $\chi^2$ function that quantifies the deviation between the measured and model-predicted intensities $q_k = \mathrm{Tr}\!\left[(\Pi_k^{(1)} \!\otimes\! \Pi_k^{(2)})\,\rho_i\right]$, where $\Pi_k^{(1,2)}$ denote the projectors for the chosen polarization settings. The density matrix $\rho_i$ is parameterized through a Cholesky decomposition, $\rho_i = A_i A_i^\dagger$, which guarantees a Hermitian, positive-semidefinite solution. After optimization, each reconstructed matrix is normalized by its trace, yielding a physically valid two-photon density matrix for every pixel across the SPAD image.

Figure~\ref{fig:densitymatrix}(a) shows the reconstructed real part of the density-matrix hologram in the linear polarization basis $\{\ket{HH},\ket{HV},\ket{VH},\ket{VV}\}$. Unlike conventional density matrices, where each entry is a scalar, here each element corresponds to a spatially resolved holographic pattern. Collectively, we refer to this representation as the density-matrix hologram. Several characteristic features can be identified with high intensities. The diagonal elements $(1,1)$ and $(4,4)$ display only the ``$\times$'' sub-pattern, which also appears in the off-diagonal elements $(1,4)$ and $(4,1)$ with comparable magnitudes. This structure reflects the fact that only the Bell states $\ket{\Psi^+}$ and $\ket{\Phi^+}$ have nonzero contributions from $\ket{HH}$ and $\ket{VV}$, with $\ket{\Psi^+}$ associated with the ``$\times$'' sub-pattern and $\ket{\Phi^+}$ corresponding to blank information. The entries $(2,2)$, $(3,3)$, $(2,3)$, and $(3,2)$ contain the ``$+$'' and ``$=$'' sub-patterns with high intensities, For the ``$+$'' sub-pattern, the $(2,3)$ and $(3,2)$ have similar magnitude but opposite signs to those in $(2,2)$ and $(3,3)$, consistent with the contributions of the Bell states $\ket{\Psi^-}$ and $\ket{\Phi^-}$, which are mapped to these two symbols. All other off-diagonal elements are expected to vanish. This block structure demonstrates that holographic information is distributed across correlated sets of matrix elements. In the present case, the reconstructed state corresponds to a nearly pure Bell-state hologram, which theoretically has a rank-1 density matrix. More generally, the density-matrix hologram can also represent mixed states of arbitrary rank, thereby encoding more complex quantum holographic information. 

To render the encoded information more directly, we project the reconstructed state onto the polarization Bell basis using the scalar map $\langle \psi_B | \rho(x,y) | \psi_B \rangle$, where $|\psi_B\rangle$ is one of the $|\Psi^+\rangle$, $|\Psi^-\rangle$, $|\Phi^+\rangle$, $|\Phi^-\rangle$. The resulting Bell-projection images are shown in Fig.~\ref{fig:densitymatrix}(b): the ``$\times$'', ``$+$'', and ``$=$'' patterns appear in the $\ket{\Psi^+}$, $\ket{\Psi^-}$, and $\ket{\Phi^-}$ projections, respectively, while the $\ket{\Phi^+}$ projection remains blank—consistent with the target Bell-state hologram in Eq.~\eqref{eq:twoPhotonHologram}. These results demonstrate that the metasurface encodes distinct holographic symbols into specific Bell components of the two-photon state, providing a basis- and spatially-resolved readout that goes beyond conventional two-qubit tomography~\cite{Erhard2020NatRevPhys,Cozzolino2019AQT}. 

Once reconstructed, the density matrix provides access to quantitative properties of the holographic quantum state. Here, we grouped the reconstructed pixels according to their corresponding target Bell states and averaged the density matrices within each group. The averaged density matrices were then used to extract information about the properties of the holographic quantum states in the three regions of the holographic symbols ``$\times$'', ``$+$'', and ``$=$''. As an example, the fidelities (ranging from 0 to 1),
\(
F(\rho, \sigma) =
\left[\mathrm{Tr}\!\left(
\sqrt{\sqrt{\rho}\,\sigma\,\sqrt{\rho}}
\right)\right]^2 ,
\)
with respect to the three corresponding target states 
($\sigma = |\Psi^+\rangle\langle\Psi^+|$, $|\Psi^-\rangle\langle\Psi^-|$, and $|\Phi^-\rangle\langle\Phi^-|$)
are 0.70, 0.72, and 0.75, confirming a high degree of agreement with the intended Bell-state holograms. This high fidelity is equivalently manifested as the bright diagonal elements shown in Fig.~\ref{fig:densitymatrix}(b).

For the entanglement properties, the CHSH--Bell parameters evaluated from the density matrices are 
$S = 1.93$, $1.89$, and $2.07$ for the three groups of signal pixels, revealing strong two-photon correlations that approach the classical Bell bound ($S = 2$) but do not yet exhibit a clear nonlocal entanglement violation. In contrast, the background pixels yield $S = 0.11$ for their averaged density matrix, characteristic of a highly mixed, uncorrelated two-photon state. 
To further quantify entanglement, we evaluate the concurrence margin $M$, which is $-0.46$ for the background (negative for separable states) and $M = 0.48$, $0.47$, and $0.54$ for the three signal groups, corresponding to moderately entangled states. The reduction in $S$ primarily arises from the limited signal-to-noise ratio of the SPAD camera—whose background noise is significantly higher than that of the SPCM—rather than from the intrinsic quality of the quantum state itself.
In a modified setup (Fig.~S7), the SPAD camera in arm~$b$ was replaced with a high-SNR SPCM, and the metasurface output was collected in the far field by a fiber-coupled detector, while photon~$a$ was detected by a second SPCM. The measurement thus focused on the central spot of the density-matrix hologram corresponding to the incident Bell state $|\Phi^-\rangle$. Using the analyzer settings $(\theta_a,\theta'_a,\theta_b,\theta'_b)=(22.5^\circ,67.5^\circ,0^\circ,45^\circ)$, corresponding to the standard CHSH configuration, we obtained $S=2.55$ —exceeding the classical bound of $2$ and approaching the quantum limit of $2\sqrt{2}$—corresponding to a visibility~\cite{Virzi2024QST} of $90.2\%$ (see \emph{Supplementary Note~7} for details). This strong CHSH violation confirms the presence of nonclassical correlations between the two photons in this region of the density-matrix hologram~\cite{Brukner2004PRL}.

\section{Discussion and Conclusion}

In this work, we have introduced and demonstrated the concept of Bell-state holograms, enabled by a polarization-multiplexed metasurface that couples polarization to different spatial wavefronts with different output polarizations. 
This capability allows distinct holographic images to be attached to different Bell states of a polarization-entangled photon pair. 
Unlike conventional polarization optics, which act solely on spin angular momentum, or orbital-angular-momentum (OAM) encoding, which is limited to a specific class of spatial modes, our approach employs holograms themselves as information carriers. 
This significantly enlarges the accessible spatial-mode space and provides a versatile platform for high-dimensional photonic encoding.

In the present device, three Bell states are assigned holographic symbols while the fourth remains blank—a restriction that can be lifted in future by employing more advanced metasurface architectures. 
For example, multilayer or bianisotropic metasurfaces capable of controlling all four Jones-matrix elements of each nanostructure could independently address all four Bell states, offering complete freedom in holographic encoding.
To fully characterize the constructed Bell-state holograms, we developed and implemented a quantum hologram tomography protocol. 
This method reconstructs the quantum hologram at the density-matrix level, revealing how holographic symbols are distributed across the two-photon state. 
The pixelwise density-matrix representation provides complete information, including both probabilities and coherences, thereby capturing the full quantum correlations underlying the holographic encoding. In the current case, the SPAD camera is indispensable for spatially resolved reconstruction of the density-matrix hologram, while the SPCM provides high-fidelity verification of entanglement. We expect in the future that SPAD cameras with improved signal-to-noise ratios will enable direct Bell-inequality tests with higher accuracy on the density-matrix hologram itself.
On the other hand, while the present experiment realizes an almost pure-state Bell hologram, the same methodology naturally extends to mixed states, establishing density-matrix holography—in a broader context than Bell-state holograms—as a general characterization and quantum-information extraction tool for quantum holograms.

A distinctive feature of the reconstructed density-matrix hologram, unlike a conventional hologram, is that it conveys information not only through constructive intensity but also through contrast. 
Because the state tomography enforces $\mathrm{Tr}\,\rho(x,y)=1$ at each pixel, the reconstructed values in the background regions—where no holographic information is encoded, such as in the diagonal panels of Figs.~\ref{fig:densitymatrix}(a) and~\ref{fig:densitymatrix}(b)—fluctuate around $1/4$ for each diagonal element rather than zero. 
This behaviour arises directly from the normalization constraint. 
In contrast, within the holographic-symbol regions but in panels not corresponding to the assigned Bell states, the reconstructed values are close to zero—substantially lower than $1/4$. 
These low values should not be interpreted as enhanced noise robustness. 
Consequently, as seen in the upper-left panel of Fig.~\ref{fig:densitymatrix}(b), while the “$\times$” symbol appears with high intensity (as it is associated with the $|\Psi^+\rangle$ Bell state), the “$+$” and “$=$” symbols—associated with other Bell states of high intensity in their own panels—appear darker than the background, producing large contrast. 
This behaviour is intrinsic to density-matrix holography and should be taken into account when interpreting quantum holograms at the density-matrix level.
On the other hand, to evaluate the contrast between the signal pixels (corresponding to different target Bell states) and the background pixels, we rescale the probabilities to a conventional contrast range by using the relative von Neumann entropy as a ``quantum contrast'',
\[
C_q(\rho) = \tfrac{1}{2}\big[\mathrm{Tr}(\rho \log_2 \rho) + 2\big],
\]
which quantifies the information contrast relative to the maximally mixed state $I/4$. 
It ranges from 0 for a maximally mixed state to 1 for a pure state. 
The resulting contrast values are 0.42, 0.39, and 0.50 for the three regions displaying the holographic symbols “$\times$”, “$+$”, and “$=$”, respectively, compared with 0.02 for the background (where the density matrices are averaged within each region), indicating that the background pixels are nearly maximally mixed ($C_q \approx 0$).

Looking ahead, the Bell-state and density-matrix hologram framework introduced here opens several promising directions. 
By increasing the number of holographic symbols or encoded spatial modes, the dimensionality of the entangled Hilbert space can be extended far beyond two-level qubits. 
The systematic buildup of complexity—from polarization, to orbital angular momentum (OAM)~\cite{Devlin2017Science}, and ultimately to holographic encoding—offers a natural progression toward $d$-level (qudit) protocols and high-capacity, secure quantum communication, encryption and information processing ~\cite{Cerf2002PRL,Ekert1991PRL}. 
Furthermore, implementing metasurfaces in both photon arms could enable richer forms of holographic coupling and more complex hybrid entangled states, further enhancing the accessible information capacity. 
In summary, our results establish density-matrix holography as a powerful characterization technique and demonstrate Bell-state holograms as a new resource for high-dimensional quantum technologies.

\medskip
\noindent \textbf{Methods}


\medskip
\noindent \textbf{Simulation}

The silicon nanopillars were simulated using the finite element method, with periodic boundary conditions applied in the x and y directions, and perfectly matched layers (PML) implemented along the z direction. Circularly polarized light was used as the excitation source propagating along the z-axis. Full details are provided in \emph{Supplementary Note 3}.

\medskip
\noindent \textbf{Sample fabrication}

The sample was fabricated on a 500 $\mu$m thick quartz substrate. Initially, a layer of 500 nm $\alpha$-Si film was deposited on the quartz substrate by plasma enhanced chemical vapor deposition (PECVD), followed by spin-coating with PMMA electron beam resist of 400 nm. The desired structure was patterned using an electron beam lithography system (JEOL 6300FS) at a base dose of 1000 $\mu C/cm^2$ with an accelerating voltage of 100 kV. After exposure, the resist was developed in a MIBK/IPA solution for 40 seconds and rinsed in IPA for 30 seconds. This was followed by the deposition of an 80 nm chromium (Cr) layer using electron beam evaporation deposition. To achieve the lift-off process, the sample was immersed in hot acetone at 65°C and cleaned ultrasonically. Finally, the desired structure was transferred from Cr to silicon using inductively coupled plasma (ICP) reactive ion etching (RIE) with HBr at room temperature for 170 seconds (flow rate is 50 sccm, pressure is 10 mTorr, RF and ICP power are 50 W and 750 W, respectively). The residual Cr hard mask was removed using cerium (IV) ammonium nitrate.

\medskip
\noindent \textbf{Experiment}

Prior to conducting quantum optical experiments, we have first characterized and confirmed the metasurface function shown in Eq. \eqref{eq:M_operator} by shining a 810-nm laser to the metasurface with conventional CCD collecting the far-field classical images with details in \emph{Supplementary Note 5}.
Next, for the quantum optical experiments, polarization-entangled photon pairs were generated via SPDC in a type-II BBO crystal pumped by a 405 nm laser (CrystaLaser DL-405-400). The entangled photon pairs exhibited a half-opening angle of 3 degrees \cite{Kwiat1995PRL}. The photon a was collected through a single-mode fibre and detected by a SPCM (Excelitas SPCM-800-14-FC), whose electronic output served as an external trigger to drive the SPAD camera (SPAD512S) for capturing the correlated photon b. To ensure efficient interaction between the photons in the arm b and the metasurface, a beam reduction system composed of a lens (focal length f is 150 mm) and a 10× objective was employed to confine the majority of photons within the metasurface region. This was followed by a beam expansion system, consisting of a 50× objective and a 100 mm focal length lens, to magnify the image of the metasurface. A subsequent lens (focal length 75 mm) then focused the quantum holographic pattern onto its focal plane, where the SPAD camera was positioned to record the quantum hologram images. The experimental measurement of quantum hologram tomography was implemented by inserting rotatable analysers into both the arms a and b to project the entangled photons onto specific polarization states. This process was combined with the SPAD camera’s capability to spatially resolve the photon b in a pixelated manner, thereby capturing the spatial quantum hologram information encoded in the entangled state. Each quantum holographic image was retrieved from 10,000 frames, with each frame spanning 100 ms and a 24 ns detection window per external trigger. The complete experimental configuration is detailed in \emph{Supplementary Note 6}.

\medskip
\noindent \textbf{Data availability}
The data used in this study are available from the corresponding authors upon reasonable request.

\medskip
\noindent \textbf{Code availability}
The code that supports the findings of this study are available from the corresponding authors upon reasonable request.

\medskip
\noindent \textbf{Acknowledgements}
This work was supported by the Hong Kong Research Grants Council under Grant Nos. 16304524, STG3/E-704/23N, AoE/P-502/20, and JLFS/P-603/24; by the Croucher Foundation under Grant No. CF23SC01; and by the Postdoctoral Fellowship Program of China Postdoctoral Science Fund under Grant No. GZB20240812. We would like to acknowledge the IOP-HKUST-Joint Laboratory for Wave Functional Materials Research.

\bibliography{bib}

\end{document}